%% file: 0.tex
\documentclass{LMCS} 
\input local-defs.tex

\def\doi{3 (1:2) 2007}
\lmcsheading%
{\doi}
{1--16}
{}
{}
{Sep.~23, 2005}
{Jan.~25, 2007}
{}   

\begin{document}
\title{Cores of Countably Categorical Structures}
\author{Manuel Bodirsky}
\address{Institut f\"ur Informatik, Unter den Linden 6, Humboldt-Universit\"at zu Berlin}
\email{bodirsky@informatik.hu-berlin.de}

\titlecomment{This is an extended and corrected version of a conference paper titled
``The Core of a Countably Categorical Structure''~\cite{BodCores}}


\input abstract.tex
\maketitle
\input mathintro.tex

\input definitions.tex
\input countcat.tex\input homogeneous.tex

\input pp.tex
\input main.tex
\input constants.tex

\section{Acknowledgements}
I am grateful to Julia B\"ottcher, Peter Cameron, Gregory Cherlin, 
Daniel Kr\'al, Martin Ziegler for discussions. 
Special thanks to anonymous referees,
who greatly helped with their comments.

\bibliographystyle{abbrv}
\bibliography{local}
\end{document}

%% file: local-defs.tex
\usepackage{url}
\usepackage{cite}
\usepackage{amsthm, amsmath,amsfonts,amsxtra, amssymb, enumerate}
\usepackage{eucal, url}
\usepackage{xspace, varioref}
\usepackage{hyperref}
\theoremstyle{plain}
\newtheorem{theorem}{Theorem}
\newtheorem{proposition}[theorem]{Proposition}
\newtheorem{corollary}[theorem]{Corollary}
\newtheorem{lemma}[theorem]{Lemma}
\theoremstyle{definition}
\newtheorem{definition}[theorem]{Definition}

\newcommand{\cproblem}[3]{
\vspace{.2cm}
\noindent {\bf #1} \\
INSTANCE: #2 \\
QUESTION: #3 \\}

\newcommand{\Fresse}{Fra\"{i}ss\'{e}}

\newcommand{\Th}{\text{Th}}

\newcommand{\ignore}[1]{}


%% file: abstract.tex
\begin{abstract}
A relational structure is a \emph{core}, if all its
endomorphisms are embeddings. This notion is important for
computational complexity classification of constraint satisfaction problems.
It is a fundamental fact that
every finite structure has a core, i.e.,  an endomorphism 
such that the structure induced by its image is a core;
moreover, the core is unique up to isomorphism.

We prove that every \emph{$\omega$-categorical} structure has a core.
Moreover, every $\omega$-categorical structure is homomorphically 
equivalent to a model-complete core, which is unique up to
isomorphism, and which is finite or $\omega$-categorical.
We discuss consequences for 
constraint satisfaction with $\omega$-categorical templates.
\end{abstract}

%% file: mathintro.tex
\section{Introduction}\label{sect:intro}
The notion of the \emph{core} of a finite structure is a central concept
in structural combinatorics~\cite{HNBook}. It can be defined in many equivalent
ways, one of which is as follows (for detailed definitions see Section~\ref{sect:cores}). A finite structure $S$ is a \emph{core} if every endomorphism of $S$ is an automorphism. A structure $C$ is called a \emph{core of
$S$} if $C$ is a core and is isomorphic to a substructure induced by the image
of an endomorphism of $S$.

For finite structures $S$ it is well-known and easy to prove
that a core of $S$ always exists, and 
is unique up to isomorphism; see e.g.~\cite{HNBook}. 
We therefore speak of \emph{the core of} a finite relational structure $S$.
For infinite structures
various core-like properties were studied by
Bauslaugh~\cite{InfCores, InfDigraphCores}. 
In general, infinite structures might not have a core in the sense introduced
above, see~\cite{InfCores, InfDigraphCores}.

An important application of the concept of a core is the theory of
\emph{constraint satisfaction} in computer science. 
Roughly speaking, a constraint
satisfaction problem is a computational problem where the input consists 
of a given set of variables and a set of constraints that are imposed on these variables, and where the task is to assign values to the variables such
that all the constraints are satisfied. Such problems appear in 
numerous areas of computer science, and constraint satisfaction problems
attracted considerable attention in artificial intelligence.

Many constraint satisfaction problems
can be formalized as follows, using the concept of structure homomorphisms;
again, for a detailed introduction of all the involved concepts see Section~\ref{sect:defs}. Let $\Gamma$ be a finite or infinite 
structure with relational signature $\tau$. 
The \emph{constraint satisfaction problem (CSP)} 
for the so-called \emph{template} $\Gamma$
is the following computational problem.

\cproblem{CSP($\Gamma$)}
{A finite structure $S$ of the same relational signature $\tau$ as the template $\Gamma$.}
{Is there a homomorphism $h: S \rightarrow {\Gamma}$?}

We want to stress that $\Gamma$ is not part of the input; each $\Gamma$
defines a computational problem.
Note that if two structures $\Gamma$ and $\Delta$ are 
\emph{homomorphically equivalent}, i.e., there is a homomorphism from
$\Gamma$ to $\Delta$ and from $\Delta$ to $\Gamma$, then
these structures have the same constraint satisfaction problem.
In particular, a structure $S$ and its core $C$ have the same constraint
satisfaction problem.

For a finite template $\Gamma$, the computational problem CSP$(\Gamma)$ is clearly
contained in NP. A classification of tractable 
and NP-hard constraint satisfaction problems with a finite template
is intensively studied, 
but still not complete. See
\cite{HellNesetril, FederVardi, JBK}, 
just to mention a few highlights on that subject.
In all these approaches, the authors make use of the assumption 
that the templates of the constraint satisfaction problems under consideration are cores.

The class of constraint satisfaction problems with an \emph{infinite} template
was not yet studied systematically.
It turns out that many interesting computational problems can be
formulated with templates that are \emph{$\omega$-categorical}.
The concept of $\omega$-categoricity is central in classical model theory,
and will be introduced carefully in Section~\ref{sect:cat}.
The following list of well-known computational problems
can all be formulated with a countably infinite template that is
$\omega$-categorical.

\begin{enumerate}[$\bullet$]
\item Allen's interval algebra, and all its fragments \cite{Allen, Nebel, KrokhinAllen, Hirsch}
\item Problems in phylogenetic analysis \cite{Steel, Gusfield}
\item Tree description constraints in computational linguistics 
\cite{Cornell, BodirskyKutz, BodirskyNesetril}
\item Computational problems in the theory of relation 
algebras~\cite{HirschAlgebraicLogic,LadkinMaddux,Duentsch}
\item All CSPs in monotone monadic SNP without inequality 
\cite{FederVardi, BodirskyDalmau}
\end{enumerate}

Moreover, 
every constraint satisfaction problem with a finite template $\Delta$
can also be formulated with an $\omega$-categorical template.

In this article we study how the notion of the core of a finite relational
structure can be generalized to $\omega$-categorical structures.
We say that an infinite structure $\Gamma$ is a \emph{core}, if 
every endomorphism $e$ of $\Gamma$ is an embedding, i.e., $e$ is an
isomorphism between $\Gamma$ and the structure induced by the image of $e$
in $\Gamma$. This is indeed a generalization of the previous definition
for finite structures, since every embedding of a finite structure $\Delta$ in $\Delta$
is an automorphism. As in the finite case, 
we say that \emph{$\Gamma$ has a core $\Delta$}, if $\Delta$ is 
a core and $\Delta$ is isomorphic to the structure induced in $\Gamma$ by the
image of some endomorphism of $\Gamma$.

Note that our definition of a core is fundamentally different from
the notion of a \emph{core structure} as introduced in~\cite{Kueker} 
in model theory.
There, a structure $\Gamma$ is called a 
\emph{core structure of a first-order theory $T$} if $\Gamma$ is isomorphic to 
exactly one substructure of every model of $T$.

\paragraph{Results.} We will show that 
every $\omega$-categorical structure $\Gamma$ has a core.
Moreover, every $\omega$-categorical $\tau$-structure $\Gamma$ 
is homomorphically equivalent to a model-complete core $\Gamma^c$,
which is unique up to isomorphism, and which is finite
or again $\omega$-categorical. 

If $\Gamma$ is expanded by all primitive positive definable
relations, then $\Gamma$ is homomorphically equivalent to
a homogeneous core $\Gamma^c$, which is unique up to isomorphism, 
and which is finite or $\omega$-categorical.
The condition that $\Gamma$ is expanded by all primitive positive
definable relations is natural in the context of constraint satisfaction,
since a relational structure and its expansion by all primitive
positive definable relations have the same computational complexity 
(see e.g.~\cite{JeavonsClosure}). Since $\Gamma^c$ is in this case
homogeneous, it admits quantified elimination and has a
$\forall\exists$-axiomatization.

In Section~\ref{sect:constants} we also prove the following 
result, which has consequences for the theory of constraint satisfaction.
If we expand $\Gamma^c$ by a singleton relation, 
then the resulting constraint satisfaction problem
has the same computational complexity as CSP$(\Gamma^c)$.
This was shown for finite templates in~\cite{JBK},
and is of fundamental importance for the so-called 
\emph{algebraic approach} to constraint satisfaction.

%% file: definitions.tex
\section{Cores}\label{sect:defs}
\label{sect:cores}

Let $\Gamma$ and $\Delta$ be relational structures with the same relational 
signature $\tau$. We use the same symbols for the relation 
symbols from $\tau$ and for the respective relations in $\Gamma$,
and the same symbols to denote a structure and its universe.
A mapping $f: \Gamma \rightarrow \Delta$ is called a \emph{homomorphism}, 
if for all relation symbols $R \in \tau$ 
and $x_1, \dots, x_n \in \Gamma$ the relation
$R(f(x_1), \dots, f(x_n))$ holds in $\Delta$ 
whenever $R(x_1, \dots, x_n)$ holds in $\Gamma$.
A homomorphism is called \emph{strong}, 
if $R(x_1, \dots,$ $x_n)$ holds in $\Gamma$ 
\emph{if and only if} $R(f(x_1), \dots, f(x_n))$ holds in $\Delta$.
An injective strong homomorphism from $\Gamma$ to $\Delta$
is also called an \emph{embedding}; if there exists an embedding of
$\Gamma$ into $\Delta$, then $\Delta$ is called an \emph{extension}
of $\Gamma$. A homomorphism
from $\Gamma$ to $\Gamma$ is called an \emph{endomorphism} of $\Gamma$,
and a bijective strong endomorphism of $\Gamma$ is called an 
\emph{automorphism} of $\Gamma$.
Two structures $\Gamma$ and $\Delta$
are called \emph{homomorphically equivalent},
if there is a homomorphism from $\Gamma$ to $\Delta$ and a
homomorphism from $\Delta$ to $\Gamma$.

\begin{definition}\label{def:core}
A (finite or infinite) structure $\Gamma$ is a \emph{core}
if every endomorphisms of $\Gamma$ is an embedding.
A core $\Gamma$ is called \emph{a core of $\Delta$} 
if $\Gamma$ is the image of an endomorphism of $\Delta$. 
\end{definition}

Homomorphisms that are not embeddings are called \emph{strict}.
The above definition says that cores do not have strict endomorphisms.
For \emph{finite} cores it clearly holds that every endomorphism 
is an automorphism, hence our definition is a 
generalization
of the notion of a core for finite structures.


The following proposition illustrates the relevance of cores
for constraint satisfaction problems.

\begin{proposition}
Let $\Gamma$ be a relational structure.
Then CSP$(\Gamma)$ can be formulated as a constraint satisfaction problem 
with a finite
template if and only if $\Gamma$ has a finite core.
\end{proposition}
\begin{proof}
Clearly, if $\Gamma$ has a finite core $\Delta$, then CSP$(\Gamma)$ is equivalent
to CSP$(T)$. Conversely, suppose CSP$(\Gamma)$ equals CSP$(\Delta)$ for
a finite template $\Delta$. This implies that 
every finite substructure of $\Gamma$
homomorphically maps to $\Delta$. A standard compactness argument shows
that there is a homomorphism from $\Gamma$ to $\Delta$, and therefore $\Gamma$ has a finite core.
\end{proof}


%% file: countcat.tex
\section{Countably Categorical Structures}\label{sect:cat}

Finite structures are up to isomorphism determined by their first-order theory.
We can not expect this for infinite structures: by the theorem of L\"owenheim-Skolem, every consistent theory with a model of cardinality $\lambda$
has models of arbitrary cardinality $\geq \lambda$.
However, it might still be the case that all models of a certain cardinality are isomorphic. If this is the case for the countably infinite models, we call the theory \emph{$\omega$-categorical}. A countably infinite 
structure $\Gamma$ is called \emph{$\omega$-categorical}, 
if its first-order theory \emph{Th}$(\Gamma)$
(i.e., the set of all first-order sentences that hold in $\Gamma$, 
where the atomic formulas are built from the symbols in $\tau$ and equality)
is $\omega$-categorical. Throughout the paper we only consider relational 
and at most countable structures and signatures. 
Despite the powerful theorems quoted below, 
the class of $\omega$-categorical structures remains somewhat mysterious, 
and all classification results require some additional properties 
(stability in e.g.~\cite{LachlanSurvey}, or homogeneity in~\cite{Cherlin}). 

Let $G$ be a permutation group, and let $G$ act on $X$.
For all $k\geq 1$ there is a natural action of $G$ on the set $X^k$ of $k$-tuples over $X$, defined by $(x_1,\dots,x_k)\pi = (x_1 \pi,\dots,x_k \pi)$ for permutations $\pi$ from $G$.
An orbit of $k$-tuples in $G$ (with respect to this action) 
is a smallest subset $S$ of $X^k$ such 
that $\overline x \in S$ implies that $\overline x \pi \in S$ for
all $\pi \in G$.
All notions used here are standard and can be found 
e.g. in~\cite{Hodges}.

\begin{theorem}[Engeler, Ryll-Nardzewski, Svenonius]\label{thm:ryll}
The following properties of a countably infinite 
structure $\Gamma$ are equivalent:
\begin{enumerate}[\em(1)]
\item the structure $\Gamma$ is $\omega$-categorical;
\item for each $n\geq 1$, the automorphism group of $\Gamma$ contains finitely many orbits of $n$-tuples;
\item for each $n\geq 1$, $\Gamma$ admits finitely many inequivalent formulas with $n$ free variables.\qed
\end{enumerate}
\end{theorem}
Permutation groups with the second property in Theorem~\ref{thm:ryll} are 
called \emph{oligomorphic}~\cite{Oligo}. 
A famous example of an $\omega$-categorical structure is ${(\mathbb Q,<)}$,
the dense linear order of the rational numbers. 
A famous structure that is not $\omega$-categorical is $(\mathbb N,<)$. 
This structure has an infinite number of orbits of pairs, 
and hence can not be $\omega$-categorical,
because whenever $x_1-x_2 \neq y_1-y_2$, then $(x_1,x_2)$ can not be in the 
same orbit as $(y_1,y_2)$.

Constraint satisfaction with $\omega$-categorical templates is a
strict extension of constraint satisfaction with finite templates.
Let $\Delta$ be a finite template for a constraint satisfaction problem.
To formulate CSP$(\Delta)$ with an $\omega$-categorical template, 
add for each vertex $v$ in $\Delta$ a countably
infinite number of copies $v_1, v_2, \dots$, such that for all $i \geq 1$ 
the relation $R(\dots, v_i, \dots)$ holds in the resulting structure $\Gamma$
if and only if $R(\dots, v, \dots)$ holds in $\Delta$.  
It is not hard to see that the structure $\Gamma$ is $\omega$-categorical,
and that the core of $\Gamma$ is isomorphic to the core of $\Delta$.
Clearly, there are constraint satisfaction problems with $\omega$-categorical templates that can not be formulated with finite templates: all the 
classes of computational problems mentioned in the introduction 
contain examples of such problems.

To illustrate the concepts we have seen so far,
we formulate several well-known computational
problems as constraint satisfaction problems with $\omega$-categorical templates. In all these examples, 
it is fairly easy to check that the chosen template 
is a core. 
We also do not always prove $\omega$-categoricity of these structures,
and that they indeed have the specified constraint satisfaction problem.
But references with further illustration and proofs will be given.
Three more examples follow at the end of Section~\ref{sect:cat},
and for these examples the verification that the structures are indeed 
cores is more interesting. We do not present these examples here, 
because we need the concept of amalgamation 
to define them conveniently, 
which will be introduced in Section~\ref{sect:hom}.

\paragraph{Betweenness.} The hardness of many fragments of Allen's Interval Algebra~\cite{Allen, KrokhinAllen} 
can be proven easily by reduction from Betweenness, an important NP-hard problem that can be found in Garey and Johnson~\cite{GareyJohnson}.
Given a finite set $V$, and a collection $C$ of ordered triples $(x,y,z)$ of distinct elements from $V$, the computational question is whether
there is an injective function $f: V \rightarrow \{1, \dots, |V|\}$ such that, for each $(a,b,c)\in V$, we have either $f(a)<f(b)<f(c)$ or $f(c)<f(b)<f(a)$.
The formulation as a constraint satisfaction problem is straightforward,
using for instance the rational numbers as the base set of the template.

\paragraph{Switching-Acyclicity.} 
Given a digraph $D = (V;E)$, can we partition the vertices $V$ into two parts, such that the graph that arises from $D$ by switching all arcs between the two parts is acyclic? 
To formulate this as a constraint satisfaction problem
with an $\omega$-categorical template,
consider a dense subset $X$ of $\mathbb Q$, and switch the order $<$ between
the elements of $X$ and $\mathbb Q - X$, and leave the edges within $X$ and within $\mathbb Q - X$ unchanged. 
The resulting structure is called $S(2)$ and is
isomorphic for all choices of dense sets $X$, see e.g.~\cite{Cherlin}.
The constraint satisfaction problem of $S(2)$ is the problem described above~\cite{BodirskyNesetril}. For equivalent definitions of $S(2)$ and an hardness-proof of its
constraint satisfaction problem, see~\cite{BodirskyNesetril, Bodirsky}.

\paragraph{Partial tree descriptions.} Our next example is a computational
problem that was studied in 
computational linguistics~\cite{Cornell}. A polynomial time algorithm can be found in~\cite{BodirskyKutz}.
Let $D$ be a digraph with two types of arcs, called \emph{ancestorship} and \emph{non-ancestorship} arcs.
The question is whether $D$ is a 
\emph{consistent partial tree description}, i.e.,
whether we can find a forest with oriented edges on the vertex set of $D$, 
such that for every ancestor arc in $D$ 
there is a directed path in the forest, 
and for every non-ancestor arc there is no directed path in the forest.
As shown in~\cite{BodirskyNesetril}, we can 
formulate this problem as a constraint satisfaction problem
with an $\omega$-categorical template.


\paragraph{Non-cores.}
Of course, there are plenty of $\omega$-categorical structures
that are not cores, for instance the Random graph 
{\bf R}~\cite{Oligo, Hodges}, 
whose core is the complete graph $K_\omega$ on a countably infinite set of
vertices (the constraint satisfaction problem of ${\bf R}$ 
and $K_\omega$ is trivial). 


%% file: homogeneous.tex
\section{Homogeneous Structures and Amalgamation Classes}
\label{sect:hom}
We need another concept, which is of a more combinatorial nature, 
and links $\omega$-categoricity via \emph{homogeneity} and \Fresse's theorem 
to \emph{amalgamation classes}. A structure is \emph{homogeneous} 
(sometimes also called \emph{ultra-homogeneous}~\cite{Hodges})
if every isomorphism between
finite substructures of $\Gamma$ can be extended to an automorphism
(in this paper, substructure always means \emph{induced substructure},
as in~\cite{Hodges}).
A structure $\Gamma$ admits \emph{quantifier elimination}, 
if every first-order
formula has in $\Gamma$ a quantifier-free definition. 
\begin{proposition}[see e.g.~2.22 in~\cite{Oligo}, and~\cite{Hodges}]
\label{prop:qe}
An $\omega$-categorical structure has quantifier elimination if and only
if it is homogeneous. A countable 
homogeneous structure $\Gamma$ is $\omega$-categorical 
if $\Gamma$ contains finitely many relations of arity
$k$, for all $k \geq 1$.\qed
\end{proposition}

For an example of a homogeneous structure that is not $\omega$-categorical, consider the expansion of a countably infinite structure 
$\Gamma$ by unary singleton predicates for each element in $\Gamma$.
This structure is homogeneous, since there are no distinct isomorphic
substructures in $\Gamma$,
and it is not $\omega$-categorical, since the number of orbits in the automorphism group of $\Gamma$ is infinite.

The next theorem asserts that 
a countable homogeneous structure is up to isomorphism characterized by 
its \emph{age}. The \emph{age} of a relational structure $\Gamma$
is the set of finite structures that embed into $\Gamma$ (this
is terminology that goes back to \Fresse~\cite{Fraisse}).
A class of finite relational structures ${\mathcal C}$ is an \emph{amalgamation class} if ${\mathcal C}$ is nonempty, closed under isomorphisms and taking substructures, and has the \emph{amalgamation property}, which says that for all $A,B_1,B_2 \in {\mathcal C}$ and embeddings $e_1: A \rightarrow B_1$ and $e_2: A \rightarrow B_2$ there exists $C \in {\mathcal C}$ and embeddings $f_1: B_1 \rightarrow C$ and $f_2: B_2 \rightarrow C$ such that $f_1e_1=f_2e_2$. 
\begin{theorem}[\Fresse~\cite{Fraisse}]\label{thm:fraisse}
A countable class ${\mathcal C}$ of finite relational structures with countable signature is the age of a countable homogeneous structure if and only if ${\mathcal C}$ is an amalgamation class. In this case the homogeneous structure
is up to isomorphism unique and called the \emph{Fra\"\i ss\'e-limit} 
of $\mathcal C$.\qed
\end{theorem}
The following templates of well-known constraint satisfaction problems
are easily defined with amalgamation classes.

\paragraph{Triangle-freeness.} 
Given a graph $G$, is $G$ triangle-free? Clearly, this problem can be
solved in polynomial time.
However, it can not be formulated as a constraint satisfaction problem 
with a \emph{finite} template.
To formulate this problem as a constraint satisfaction problem with
an $\omega$-categorical template, note that
the class of all triangle-free graphs is an amalgamation class.
Let us denote its \Fresse-limit by $\ntriangleleft$.
It is well-known and not hard to see~\cite{Oligo} 
that this graph is up to isomorphism 
uniquely determined by the fact that it is triangle-free and has 
the following extension property: for all finite 
subsets $A,B$ of vertices of $\ntriangleleft$ such that all vertices in $A$
are pairwise not adjacent there exists a vertex $z$ in $\ntriangleleft$
that is not in $A$, not in $B$, adjacent to all vertices in $A$, and not adjacent to all vertices in $B$. 
Clearly, CSP$(\ntriangleleft)$ is the computational problem described above. 

We claim that the structure $\ntriangleleft$ is a core.
Suppose otherwise that there is a strict endomorphism $e$. 
If $e(u)=e(v)$, then $u$ and $v$ can not be connected in $\ntriangleleft$.
We apply the extension property twice to derive that
there must be adjacent vertices $w$ and $w'$ in $\ntriangleleft$
such that $w'$ is connected to $u$ and $w$ is connected to $v$. But
then, $e(u)=e(v),e(w),e(w')$ form a triangle, a contradiction.
Hence, since $e$ is strict, there must be non-adjacent vertices $a$ and $b$
such that $e(a)$ is adjacent to $e(b)$. Again by the extension property 
$\ntriangleleft$
contains a vertex $w$ that is adjacent to both $a$ and $b$. But then 
$e(a),e(w),e(b)$ form a triangle, again a contradiction. Therefore
every endomorphism of $\ntriangleleft$ is an embedding.

\paragraph{No-mono-tri.}
The structure $[\ntriangleleft, \ntriangleleft]$, i.e., the 
structure that consists of two copies of $\ntriangleleft$,
where all vertices between the two copies are linked, has an interesting 
constraint satisfaction problem, which can be formulated as follows:
Given a graph, can we partition its vertices into two parts such that both 
parts do not contain a triangle? This problem is a rather typical example 
from the class \emph{monotone monadic SNP without inequality (MMSNP)},
a fragment of existential second-order logic introduced in~\cite{FederVardi}
in the context of constraint satisfaction. 
A general result on so-called \emph{$G$-free colorability}
implies its NP-hardness~\cite{Achlioptas}.
Every constraint
satisfaction problem in MMSNP can be formulated with an $\omega$-categorical
template~\cite{BodirskyDalmau}. The construction given in~\cite{BodirskyDalmau} also shows that
the above structure $[\ntriangleleft, \ntriangleleft]$ is $\omega$-categorical.
Similarly as in the previous example for the graph $\ntriangleleft$,
it is not hard to show that $[\ntriangleleft, \ntriangleleft]$ is a core.

\paragraph{Quartet compatibility.} 
The next example is an important structure in the theory of infinite
permutation groups\cite{Oligo}. A \emph{boron tree} is a finite tree 
in which all vertices have degree one (\emph{hydrogen atoms}) or
degree three (\emph{boron atoms}). On the hydrogen atoms of
a boron tree we can define a quaternary relation $xy|uv$ that holds when the paths joining $x$ to $y$ and $u$ to $v$ are disjoint.
The class of all structures $\mathcal D$ with a quaternary relation that stem from a boron tree as defined above is an amalgamation class~\cite{AdelekeNeumann}. 
Let $D$ be the Fra\"{i}ss\'{e}-limit of $\mathcal D$. 
Then CSP$(D)$ is a well-known NP-hard problem~\cite{Steel} that
was independently studied in phylogenetic analysis (without any reference
to constraint satisfaction), and is called \emph{quartet-compatibility}:
Given a collection $C$ of quartets $xy|uv$ over a set $X$,
is there some tree with leaf set $X$ 
such that for each quadruple $xy|uv$ in $C$ the paths from 
$x$ to $y$ and from $u$ to $v$ do not have common vertices?

\paragraph{Rooted triple consistency.} 
The next problem is studied in phylogenetic analysis, again without
notice that the problem can be stated as a constraint satisfaction problem.
If we fix a point $a$ in the previous structure $D$ 
and consider the ternary relation `:' 
defined by $x:yz \; \Leftrightarrow \; ax|yz$, we again 
obtain an $\omega$-categorical structure 
(this is a \emph{C-set} in \cite{AdelekeNeumann}). 
The age of this structure now contains the finite structures 
$T$ that come from finite \emph{rooted} trees, and the relation $x:yz$ 
says that the least common ancestor of $y$ and $z$ is strictly below the least 
common ancestor of $x,y,$ and $z$ in the tree $T$. 
The corresponding constraint satisfaction problem is known
as the rooted triple consistency problem~\cite{Steel}, and tractable.
The first polynomial time algorithm for this problem 
goes back to~\cite{AhoUllman}, motivated by a question
in database theory.

%% file: pp.tex
\section{Existential Positive Expansions}
In this section we study various syntactic restrictions
of first-order logic. Recall that 
if relations are added to a given $\tau$-structure $\Gamma$ then
the resulting structure $\Gamma'$ is called an \emph{expansion} of $\Gamma$, 
and $\Gamma$ is called the \emph{$\tau$-reduct} of $\Gamma'$. 

A first-order formula $\phi$ is \emph{primitive (primitive positive)}, 
if it is of the form 
\begin{align*}
\exists \overline{x}. \psi_1 \wedge \dots \wedge \psi_k \,
\end{align*}
where $\psi_i$ are literals (atomic formulas) that might include the
equality relation. It is called 
\emph{existential (existential positive)}, if 
it is of the form $\exists \overline{x}. \Psi$ where $\Psi$ is quantifier-free
(and negation-free). 
The strongest of these four syntactic
restrictions, \emph{primitive positivity}, is important for constraint
satisfaction, since the expansion of a template with primitive positive
definable (short, \emph{pp-definable}) 
relations does not change the complexity of the corresponding constraint satisfaction problem. This is an easy observation, 
see e.g.~\cite{JeavonsClosure}.

A mapping of a $\tau$-structure $\Gamma$ to a $\tau$-structure $\Delta$ is
called \emph{elementary}, if it preserves all first-order $\tau$-formulas 
(this is a standard notion in model-theory \cite{Hodges}).
A theory $T$ is called \emph{model-complete}, if all embeddings
between models of $T$ are elementary.
In the case that $T$ is the theory of a structure
$\Gamma$, we also say that $\Gamma$ is model-complete, as usual.
If $T$ is a first-order theory, then $\Gamma$ is called an
\emph{existentially complete} model of $T$ if for every existential
formula $\phi$, every tuple $\overline a$ in $\Gamma$, and every
model $\Delta$ of $T$ such that $\Gamma$ is a substructure of $\Delta$
and $\phi(\overline a)$ holds in $\Delta$, the formula
$\phi(\overline a)$ also holds in $\Gamma$.

Model-completeness can be characterized in many different ways.
We say that two formulas $\phi,\psi$ are \emph{equivalent with respect to} a
theory $T$ if in every model of $T$ the formulas $\phi$ and $\psi$ define
the same relations.

\begin{proposition}[Theorem 7.3.1 in~\cite{Hodges}]\label{prop:mc}
Let $T$ be a first-order theory. Then the following are equivalent.
\begin{enumerate}[$\bullet$]
\item $T$ is model-complete
\item Every model of $T$ is an existentially complete model of $T$ 
\item Every first-order formula is equivalent to an existential formula
with respect to $T$.\qed
\end{enumerate}
\end{proposition}

An $\omega$-categorical structure is model-complete if and only if 
its first-order theory is equivalent
to a set of \emph{$\forall \exists$-sentences}, i.e., 
sentences of the form $\forall \overline x \exists \overline y
\phi(\overline x, \overline y)$, where $\phi$ is quantifier-free
(see e.g.~Theorem 7.3.3 and 7.3.4 in~\cite{Hodges}).
The following proposition 
follows directly from Theorem~7.2.1 in~\cite{Hodges}.

\begin{proposition}\label{prop:ec}
Let $\Gamma$ be a relational structure, and let $T$ be the
set of all $\forall\exists$-sentences that hold in $\Gamma$.
Then there exists an extension of $\Gamma$ that is 
an existentially complete model of $T$.\qed
\end{proposition}

The case that all existential formulas are equivalent to existential positive 
formulas can be characterized in a different way.

\begin{lemma}\label{lem:exposex}
Let $T$ be a first-order theory such that every homomorphism between models
of $T$ is an embedding. Then every existential formula
is equivalent to an existential positive formula with respect to $T$.
\end{lemma}

\begin{proof}
A formula $\phi$ 
is equivalent to an existential positive formula with respect to
a theory $T$ if $\phi$ is preserved 
by all homomorphisms\footnote{We say that a formula $\phi$ is \emph{preserved 
by all homomorphisms between models of $T$} if whenever $h$ is a homomorphism from $M_1$ to $M_2$ that maps a tuple $\overline a$ of elements from $M_1$ pointwise to a tuple $\overline b$ of elements from $M_2$, where $M_1$ and $M_2$ are models of $T$, and $\overline a$ satisfies $\phi$ in $M_1$, then $\overline b$ satisfies $\phi$ in $M_2$.}
 between models of $T$;
this fact is stated as Exercise~2 in Section 5.5 in~\cite{Hodges}.
Let $f$ be a homomorphism
between two models of $T$. By assumption, $f$ is an embedding, and therefore clearly preserves all existential formulas. Hence, all
existential formulas are equivalent to an existential positive formula with respect to $T$.
\end{proof}

The following lemma will be useful to construct homogeneous models.

\begin{lemma} \label{lem:primitiveamalg}
If $\Gamma$ is a structure that has been expanded by all primitive definable
relations, then there is a homogeneous structure with the same age
as $\Gamma$.
\end{lemma}

\begin{proof}
Let ${\overline a} = (a_1, \dots, a_k)$ be a tuple of elements 
from $\Gamma$, let $B_1,B_2$ be finite induced substructures of 
$\Gamma$ and $e_1: \overline a \rightarrow B_1$ and 
$e_2: \overline a \rightarrow B_2$ be embeddings. 
Since there are relation symbols for every primitive formula in the 
signature, there is a relation $R_1$ that holds on the tuple 
$e_1(\overline a)$ in $\Gamma$, and is defined by 
the following primitive formula 
$\phi$. Let $b_1,\dots,b_m$ be the elements of $B_1$.
The formula $\phi$ has $k$ free variables $x_1,\dots,x_k$ 
and has the form $\phi := \exists x_{k+1}, \dots, x_{m} \psi$.
The formula $\psi$ is a conjunction of literals defined as follows.
If there is an $l$-ary relation symbol $R$ in the signature of $\Gamma$
such that $R$ holds on elements $b_{i_1},\dots,b_{i_l}$ in $B$, 
$1 \leq i_1,\dots,i_l \leq m$, 
then $\psi$ contains the conjunct $R(x_{i_1},\dots,x_{i_l})$.
If the relation $R$ does not hold on $b_{i_1},\dots,b_{i_l}$ in $B$, 
then $\psi$ contains the conjunct $\neg R(x_{i_1},\dots,x_{i_l})$.
Moreover, $\psi$ contains conjuncts of the form $x_i \neq x_j$ for all 
distinct indices $i,j$ from $1,\dots, m$.

We also have a relation $R_2$ corresponding in an analogous way to $B_2$
 where the points from $B_2-e_2({\overline a})$ are existentially quantified,
and which holds on $e_2(\overline a)$ in $\Gamma$.
Since $e_1$ and $e_2$ are embeddings, 
these relations also hold on $\overline a$ in $\Gamma$.
They assert that we can find an extension $C$ of 
the structure induced by $\overline a$ and
embeddings $f_1: B_1 \rightarrow C$, 
$f_2: B_2 \rightarrow C$ such that $f_1 e_1 = f_2 e_2$.
Thus, the age of $\Gamma$ has the amalgamation
property, and Theorem~\ref{thm:fraisse} implies that 
there is a homogeneous structure with the same age.
\end{proof}

We can combine Lemma~\ref{lem:exposex}
and Lemma~\ref{lem:primitiveamalg} and obtain the following.

\begin{corollary}\label{cor:hom}
Let $\Gamma$ be a structure that has been expanded by all 
existential positive definable relations.
If all homomorphisms between models of \emph{Th}$(\Gamma)$ are embeddings, 
then there is a homogeneous structure with the same age as $\Gamma$.
\end{corollary}

\begin{proof}
Proposition~\ref{lem:exposex} shows that every existential, and in particular
every primitive formula is in $\Gamma$ equivalent to an 
existential positive formula. Hence, $\Gamma$ is also expanded by
all primitive definable relations, and Proposition~\ref{lem:primitiveamalg}
shows that there is a homogeneous structure that has the same age
as $\Gamma$.
\end{proof}


\ignore{
\begin{proposition}\label{prop:qe}
Let $\Gamma$ be an $\omega$-categorical structure 
that has been expanded by all primitive positive definable relations.
If all endomorphisms of $\Gamma$ are embeddings, 
then there is a homogeneous structure with the same age as $\Gamma$.
\end{proposition}

\begin{proof}
Proposition shows~\ref{prop:exposex} that every existential, and in particular
every primitive formula is in $\Gamma$ equivalent to an 
existential positive formula. Therefore, the expansion $\Gamma'$ of $\Gamma$
by all existential positive definable relations satisfies the conditions
of Proposition~\ref{prop:primitiveamalg}, and there is 
a homogeneous structure that has the same age as $\Gamma'$.
Since there is a finite number of $k$-ary relation symbols in $\tau$,

Proposition~\ref{prop:qe} shows that every relation in $\Gamma'$
is a Boolean combination
of 

Hence, $\Gamma$ is also expanded by
all primitive definable relations, and Proposition~\ref{prop:primitiveamalg}
shows that there is a homogeneous structure that has the same age
as $\Gamma$.
\end{proof}
}

%% file: main.tex
\section{Cores of Countably Categorical Structures}
\label{sect:main}
In this section we state and prove the main results of the paper.
We start with a proposition on the existence of a `youngest' endomorphic image of 
an $\omega$-categorical structure. The proof employs
a typical technique for $\omega$-categorical structures.

\begin{proposition}\label{prop:precore}
Let $\Gamma$ be an $\omega$-categorical relational $\tau$-structure.
Then there exists an endomorphism $c$ of $\Gamma$ 
such that for every
other endomorphism $g$, all finite
substructures of $c(\Gamma)$ embed into $g(\Gamma)$. This is,
there exists an endomorphic image of $\Gamma$ of smallest age.
\end{proposition}

\begin{proof}
Let $\mathcal S$ be the set of all finite $\tau$-structures $S$ such that there
is an endomorphism $g$ of $\Gamma$ so that $S$ does not embed into $g(\Gamma)$.
We have to show that there is an endomorphism $c$ such that no structure
from $\mathcal S$ embeds into $c(\Gamma)$.
For the construction of $c$ we consider the following tree.
Let $a_1, a_2, \dots$ be an enumeration of $\Gamma$.
The vertices on level $n$ of the tree are labeled with 
equivalence classes of 
\emph{good} homomorphisms
from the structure induced by $\{a_1, \dots, a_n\}$ to $\Gamma$.
A homomorphism $h$ is good, if no structure from $\mathcal S$ embeds into
the structure induced by 
$h(\{a_1, \dots, a_n\})$ in $\Gamma$.
Two homomorphisms $g_1$ and $g_2$ are equivalent, if there exists 
an automorphism $\alpha$ of $\Gamma$ such that $g_1= g_2 \alpha$. Clearly,
if a homomorphism is good, then all equivalent homomorphisms 
and all restrictions are also good.
A vertex $u$ on level $n+1$ in the tree is connected
to a vertex $v$ on level $n$, if
some homomorphism from $u$ is the restriction of some homomorphism
from $v$. Because of $\omega$-categoricity, the tree is finitely
branching.
We want to show that the tree has vertices on each level $n$, and 
iteratively 
construct a sequence $h_1, h_2, \dots, h_k$ of homomorphisms from $\{a_1, \dots, a_n\}$ to $\Gamma$,
where the last endomorphism $h_k$ induces 
a good homomorphism.
Initially, if no structure from $\mathcal S$ imbeds into the structure induced by 
$\{a_1, \dots, a_n\}$,
we can choose the identity as a good homomorphism.
Otherwise, there is a structure $S \in \mathcal S$ that embeds into
the structure induced on
$\{a_1, \dots, a_n\}$, and an endomorphism $e$ such that $e(\Gamma)$ 
does not contain $S$. The mapping $h_1$ restricted
to $\{a_1, \dots, a_n\}$ is a strict homomorphism, because if
it was an embedding, $S$ embeds into the structure induced by the 
image of $\{a_1, \dots, a_n\}$ under this mapping, which is by
assumption impossible.

In step $i$, if no structure in $\mathcal S$ embeds into
the structure induced by $h_i(\{a_1, \dots, a_n\})$, 
then $h_i$ is a good homomorphism, and we are again done. 
Otherwise there is an endomorphism $e$ of $\Gamma$ and a structure 
$S\in \mathcal S$ that embeds into the structure induced by
$h_i(\{a_1, \dots, a_n\})$, 
such that $S$ does not embed into $e(\Gamma)$.
We can then define a homomorphism 
$h_{i+1}: \{a_1, \dots, a_n\} \rightarrow \Gamma$ 
by $h_{i+1}(x) := e(h_i(x))$, which is again a strict homomorphism.
Since in the sequence of structures induced by
$h_1(\{a_1, \dots, a_n\})$, $h_2(\{a_1, \dots, a_n\}), \dots$ 
either the number of vertices decreases 
or the number of tuples in relations increases, 
and since $\Gamma$ is $\omega$-categorical,
the sequence has to be finite.
Hence, there exists a good homomorphism from $\{a_1, \dots, a_n\}$ to $\Gamma$,
for all $n\geq 0$.
By K\"onig's tree lemma, there exists an infinite path in the tree.
Since adjacency in the tree was defined by restriction between
homomorphisms, this path defines an endomorphism $c$ of $\Gamma$. 
By construction, no structure in $\mathcal S$ embeds into $c(\Gamma)$.
\end{proof}

In the following, $c(\Gamma)$ denotes the structure induced
by the image of the endomorphism $c$ in $\Gamma$ that was constructed 
in Proposition~\ref{prop:precore}. Note that 
Proposition~\ref{prop:precore} says 
that all cores of $\Gamma$ have the same age as $c(\Gamma)$.
We will use the following well-known lemma several times (see e.g. Section 2.6 in~\cite{Oligo}). 
It can be shown by a similar application of K\"onig's lemma
as in the proof of the previous proposition.
 
\begin{lemma}\label{lem:emb}
Let $\Gamma$ be a relational structure whose age is contained in the
age of an $\omega$-categorical structure $\Delta$. Then $\Gamma$
embeds into $\Delta$.\qed
\end{lemma}

The following lemma is a central step in our arguments.

\begin{lemma}\label{lem:main}
Let $\Gamma$ be an $\omega$-categorical $\tau$-structure
that contains all existential positive definable relations.
Then every homomorphism between two structures $\Gamma_1$ 
and $\Gamma_2$ of the same age as $c(\Gamma)$ is an embedding.
\end{lemma}


\begin{proof}
Since $\Gamma$ is $\omega$-categorcial and because
the age of $\Gamma_1$ and $\Gamma_2$ is contained in the age of $\Gamma$,
Lemma~\ref{lem:emb} implies that 
both $\Gamma_1$ and $\Gamma_2$ embed into $\Gamma$. 
For simplicity of notation, we assume that $\Gamma_1$ and $\Gamma_2$ are 
substructures of $\Gamma$.
Now suppose for contradiction that $f: \Gamma_1 \rightarrow \Gamma_2$ 
is a homomorphism that is not an embedding, this is, 
for some tuple $\overline u = (u_1, \dots, u_k)$ of
elements and some $k$-ary relation $R$ in $\Gamma_1$ 
the mapping $f$ does not preserve the formula $\neg R(\overline u)$ 
or does not preserve the formula $u_1 \neq u_2$. We will then
construct an endomorphism $h$ of $\Gamma$ 
such that $h(\Gamma)$ does not
contain a copy of the substructure $S$ induced by
$\overline u$ in $\Gamma_1$. This is a contradiction:
On the one hand $S$ is a substructure of
$c(\Gamma)$, 
since $\Gamma_1$ has the same theory and thus the same age as $c(\Gamma)$.
On the other hand, since $S$ is not a substructure of $h(\Gamma)$,
Proposition~\ref{prop:precore} says that $S$ is not a substructure of $c(\Gamma)$.

To construct this homomorphism $h$ we consider an infinite but finitely
branching tree. The vertices on level $n$ in this tree 
will be labeled by equivalence classes of \emph{good} homomorphisms
from $\{a_1, a_2, \dots, a_n\}$
to $\Gamma$, where $a_1, a_2, \dots$ is an enumeration 
of $\Gamma$.
A homomorphism $g$ on level $n$ is \emph{good}, if the structure induced by 
$g(\{a_1, \dots, a_n\})$ 
does not contain an induced copy of $S$.
Two homomorphisms $g_1$ and $g_2$ are equivalent if there exists an automorphism $\alpha$ of $\Gamma$ such that $g_1 = g_2 \alpha$. 
Adjacency is defined by restriction; this is, 
two nodes on level $n$ and $n+1$ are adjacent in the tree if there are representatives $g_1$ and $g_2$ from these nodes 
such that $g_1$ is a restriction of $g_2$.
Clearly, all restrictions of a good homomorphism are again good homomorphisms,
and all homomorphisms in an equivalence class are good, or all are not good.
By $\omega$-categoricity of $\Gamma$, the tree has only a finite number
of vertices on level $n$, and in particular it follows that
the tree is finitely branching.
The crucial step is that the tree contains vertices on every level,
i.e., there exists a good homomorphism 
$h_{n}: \{a_1, \dots, a_n\} \rightarrow \Gamma$ for each $n\geq 1$.
We show this in the following; and here we use the assumption 
that $\Gamma$ contains all existential positive definable relations.

To find $h_{n}$ for each $n\geq 1$, 
we consider a sequence $(h^i_n)_{i\geq 0}$ of homomorphisms 
to $\Gamma$,
where the domain of 
$h^0_n$ is $\{a_1, \dots, a_n\}$, and the domain of $h_n^{i+1}$
equals the image of $h^i_n$.
Hence, we can define the following composed homomorphism 
$h_n^{(i)}: \{a_1, \dots, a_n\} \rightarrow \Gamma$ by
$h_n^{(i)}(x) := h_n^i(\dots h_n^{1}(h_n^{0}(x)) \dots)$.
We now define the sequence $(h_n^i)_{i\geq 0}$. 
The homomorphism $h_n^0$ is the identity.
For $i > 0$, if the structure induced by the domain of $h_n^i$ does not 
contain an induced copy of $S$, we are done, because then
$h_n^{(i-1)}$ is a good homomorphism from $\{a_1, \dots, a_n\}$ 
to $\Gamma$.
Otherwise, there are elements $(b^i_{1}, \dots, b^i_{k})$ 
in the domain of
$h_n^i$ that induce in $\Gamma$ a structure 
isomorphic to $S$.
We now define $h^i_n(b^i_j) := f(u_j)$ for $1 \leq j \leq k$, and
want to extend this mapping to a (strict) homomorphism $h_n^i$
on the other elements $b^i_{k+1}, \dots, b^i_{m}$, $m\leq n$, in the domain of $h_n^i$.
Consider the formula $\phi := \exists x^i_{k+1}, \dots, x^i_{m} \psi$
with free variables $x^i_1, \dots, x^i_k$,
where $\psi$ is a conjunction of atomic formulas defined as follows.
We use the structure induced by $b^i_{1}, \dots, b^i_m$ in $\Gamma$
to define $\psi$. For all $j_1,\dots,j_l \leq m$, the formula $\psi$ contains
a conjunct $R(x^i_{j_1},\dots, x^i_{j_l})$ iff $R$ holds on $b^i_{j_1},\dots,b^i_{j_l}$ in $\Gamma$.
It is then clear that the formula $\phi$ holds for $b^i_1, \dots, b^i_k$.

Since $\Gamma$ contains all existential positive
definable relations, the existential positive formula $\phi$ also holds on
$u_1, \dots, u_k$, since these vertices induce the same structure as
$b^i_1, \dots, b^i_k$.
Since $f$ preserves existential positive 
formulas, $\phi$ also holds on $f(u_1), \dots, f(u_k)$ in $\Gamma$. 
We thus can find witnesses $r_{k+1}, \dots, r_{m}$ in $\Gamma$ 
for the variables $x^i_{k+1}, \dots, x^i_{m}$
of the existential quantifiers in $\phi$, and
extend $h_i$ by $h^i_n(b^i_j) := r_j$ for $k+1 \leq j \leq m$.
Then $h^i_n$ clearly is a homomorphism to $\Gamma$, which is also strict, 
because it does not preserve
some of the inequalities or negated relations that hold on 
$b^i_{1}, \dots, b^i_{k}$.
Therefore the sequence $(h_n^{(i)})_i$ 
of homomorphisms
must be finite, because $\omega$-categoricity of $\Gamma$ implies 
that there are only finitely many non-isomorphic homomorphic images 
of the structure induced by $\{a_1, \dots, a_n\}$
in $\Gamma$. Let $h_n^{(i_0)}$ be the last homomorphism in this sequence. 
By construction, 
this mapping is a good homomorphism $h_{n}$ for every $n \geq k$.

Therefore, the constructed tree contains vertices on all levels, and
K\"onig's tree lemma asserts that the tree contains an infinite path. 
Since adjacency is defined by restriction, 
this path defines an infinite endomorphism 
$h$ of $\Gamma$. The image $h(\Gamma)$ does not contain an induced copy of $S$.
This contradicts the minimality property of $c(\Gamma)$ formulated in 
Proposition~\ref{prop:precore}.
Hence, every homomorphism from $\Gamma_1$ to $\Gamma_2$ is an embedding.
\end{proof}

\begin{corollary}
Every $\omega$-categorical $\tau$-structure $\Gamma$ has a core.
\end{corollary}
\begin{proof}
Let $\Delta$ be the expansion of $\Gamma$
 by all existential positive definable relations, and let $c(\Delta)$ be
the structure induced by the endomorphism $c$ constructed in Proposition~\ref{prop:precore} for $\Delta$. Lemma~\ref{lem:main} shows in particular that 
endomorphisms of $c(\Delta)$ are cores. Let $\Gamma_0$ be the $\tau$-reduct
of $c(\Delta)$. Clearly, since $c(\Delta)$ is a core, $\Gamma_0$ 
must be a core as well. Because $c$ is also an endomorphism of $\Gamma$,
we have that $\Gamma_0$ is a core of $\Gamma$.
\end{proof}

We already mentioned that 
the core of a finite structure is unique up to isomorphism,
As we will see now, an $\omega$-categorical structure might have
non-isomorphic cores.

\paragraph{Example.}
Consider the following $\omega$-categorical structure $\Gamma$. 
Let $\mathbb Q$ be the set of rational numbers, and let $X$ be a disjoint
countably infinite set. The domain of $\Gamma$ is $\mathbb Q \cup X$,
and the signature contains two binary relation symbols $<$ and $\neq$,
and one unary relation symbol $P$. For two elements $x,y$ in $\Gamma$
the relation $x < y$ holds iff $x,y \in
\mathbb Q$ and $x$ is a strictly smaller number than $y$. The relation
$x \neq y$ holds iff $x$ and $y$ are distinct elements in $\Gamma$.
Finally, $P(x)$ holds iff $x>0$ holds. 
It is easy to verify that $\Gamma$ is $\omega$-categorical.
One way to see this is by Proposition~\ref{prop:qe}, because
the expansion of $\Gamma$ by the 
unary relation $N := \{x \in \mathbb Q \; | \; x \leq 0 \}$ 
is homogeneous.

The structure $\Gamma$ has a model-complete core, which is unique 
up to isomorphism, namely the structure $\Delta := 
(\mathbb Q^+,<,\neq,P)$, where
$\mathbb Q^+$ is the set of all positive rational numbers, where
the interpretation of $<$ and $\neq$ is as before, and where $P$ denotes
the trivial unary relation that contains all elements of the domain.
The structure $\Delta$ is indeed the image of an endomorphism $c$
of $\Gamma$, for instance the endomorphism
that maps $\mathbb Q$ in an order-preserving way to the set 
$\{x \in \mathbb Q \; | \; x > 3 \}$, and maps $X$ bijectively
onto $\{x \in \mathbb Q \; | \; 0 < x \leq 3 \}$ in an arbitrary way. 
Let $\Delta$ be the structure that is induced by the image $\mathbb Q^+$
of this endomorphism. It is easy to verify that $\Delta$ is a core,
and hence $\Delta$ is a core of $\Gamma$.
It will follow from Theorem~\ref{thm:main} below that all other model-complete
cores of $\Gamma$ are isomorphic to $\Delta$.

However, 
we claim that $(V,<,\neq,P)$ has other cores that are not isomorphic to
$\Delta$, and not model-complete. For instance, 
let $\Gamma'$ be the substructure of $\Gamma$ induced by 
$\mathbb Q^+ \setminus (1,2)$, i.e., the set of positive rational
numbers without the open interval $(1,2)$. Then $\Gamma'$ is the image 
of an endomorphism $c'$ constructed similarly as above, with
the only difference that $c'$ maps $X$ bijectively onto
$\{x \in \mathbb Q \; | \; 0 < x \leq 1$ or $2 \leq x \leq 3 \}$.
It can be verified easily that the structure $\Gamma'$ is a core.
To see that $\Gamma'$ is not model-complete, consider the restriction $d$
of the mapping $c'$ to the elements of $\Gamma'$. 
Since $\Gamma'$ is a core, the mapping $d$ is an embedding of
$\Gamma'$ into $\Gamma'$. However, it is not an elementary embedding,
since the formula $\exists z.\; x<z<y$ holds for 
$x=d(1)$ and $y=d(2)$ in $\Gamma'$, 
but does not hold for 
$x=d(1)$ and $y=d(2)$ in the structure induced by $d(\Gamma')$.\\
\\

We will show that every $\omega$-categorical structure $\Gamma$
is always homomorphically equivalent to a \emph{model-complete} core,
which is unique up to isomorphism. 
For that, we first prove a stronger result
for structures 
that are expanded by all existential positive definable relations.

\begin{proposition}\label{prop:premain}
Let $\Gamma$ be an $\omega$-categorical $\tau$-structure.
If $\Gamma$ contains all existential positive definable relations,
then $\Gamma$ is homomorphically equivalent to a homogeneous core $\Gamma^c$,
which is unique up to isomorphism.
Moreover, $\Gamma^c$ is finite or $\omega$-categorical.
\end{proposition}
\begin{proof}
Let $T$ be the set of all universal sentences that are
true in the structure $c(\Gamma)$ 
constructed in Proposition~\ref{prop:precore}.
Proposition~\ref{prop:ec} shows that there is an extension $\Delta$ of
$c(\Gamma)$ that is an existentially complete model of $T$. 
Since $\Delta$ satisfies the same universal sentences
as $c(\Gamma)$, the two structures have the same age.
Because $\Gamma$ is $\omega$-categorical,
and because the age of $\Delta$ is contained in the age of $\Gamma$,
we can apply Lemma~\ref{lem:emb} to show that $\Delta$ embeds into $\Gamma$.
Assume for simplicity of notation that $\Delta$ is a substructure
of $\Gamma$.

Let $\phi$ be an existential formula,
and let $\overline t$ be a tuple from $\Delta$. We claim
that $\phi$ holds on $\overline t$ in $\Gamma$ if and only
if $\phi$ holds on $\overline t$ in $\Delta$.
One direction is clear, because $\Delta$ embeds into $\Gamma$.
%
%
%
Now, consider the restriction $d$ of $c$ to $\Delta$. 
Because $d(\Delta)$ is a substructure of $c(\Gamma)$,
and $c(\Gamma)$ is a substructure of $\Delta$,
$d$ is a endomorphism of $\Delta$.
Because $\Delta$ and $c(\Gamma)$ have the same age, 
Lemma~\ref{lem:main} implies that $d$ is an embedding.
Hence, the structure induced by $d(\Delta)$ is isomorphic to $\Delta$.
The embedding $d$ preserves the existential formula $\phi$,
and therefore $\phi$ holds on $d(\overline t)$ in $\Delta$.
Because $\Delta$ and thus also $d(\Delta)$ are 
existentially complete models of $T$,
the formula $\phi$ also holds on $d(\overline t)$ in the
structure $d(\Delta)$ (not only in $\Delta$). Since $d$ is an embedding, 
we then know that $\phi$ holds on $\overline t$ in $\Delta$.

Therefore, because the structure $\Gamma$ is expanded by
all existential positive relations, the structure $\Delta$ is also
expanded by all existential positive relations.
All models of $\Th(\Delta)$ must have the same age
as $\Delta$ and $c(\Gamma)$, and hence Lemma~\ref{lem:main}
shows that every homomorphism between two models of $\Th(\Delta)$
is an embedding. By Corollary~\ref{cor:hom} there
is a homogeneous structure $\Gamma^c$ with the same age
as $\Delta$ and $c(\Gamma)$. 
Theorem~\ref{thm:fraisse} shows that this structure is unique
up to isomorphism. 
The structure $\Gamma^c$ might be finite. If $\Gamma^c$ is infinite,
and since $\Gamma^c$
contains only finitely many $k$-ary relations for each $k \geq 1$, 
Proposition~\ref{prop:qe} shows that 
$\Gamma^c$ is $\omega$-categorical.

To conclude the proof,
we have to show that $\Gamma^c$ is homomorphically equivalent
to $\Gamma$. 
But this is clear since $\Gamma$ homomorphically
maps to $c(\Gamma)$, which embeds into $\Gamma^c$ by Lemma~\ref{lem:emb}.
Lemma~\ref{lem:emb} also shows that $\Gamma^c$ embeds into $\Gamma$,
because it has a smaller age than $\Gamma$ and because $\Gamma$ is 
$\omega$-categorical.
\end{proof}

We can now prove one of the main results.
\begin{theorem}\label{thm:main}
Every $\omega$-categorical $\tau$-structure $\Gamma$ 
is homomorphically equivalent to a model-complete core $\Gamma^c$,
which is unique up to isomorphism. The core $\Gamma^c$ is
$\omega$-categorical or finite. All orbits of $k$-tuples in
$\Gamma^c$ are pp-definable in $\Gamma^c$.
\end{theorem}

\begin{proof}
Let $\Delta$ be the expansion of $\Gamma$ by all existential positive 
definable relations. By the previous Proposition, $\Delta$ has a
homogeneous core $\Delta^c$, which is finite or $\omega$-categorical.
By Proposition~\ref{prop:qe}, the structure $\Delta^c$ has quantifier
elimination, and therefore every first-order formula is equivalent
to a quantifier-free formula (in the expanded signature of $\Delta$
and $\Delta^c$). Moreover, $\Delta^c$ is $\omega$-categorical.
Let $\Gamma^c$ be the $\tau$-reduct of $\Delta^c$. 
It is well-known (and follows from Theorem~\ref{thm:ryll}; see~\cite{Hodges}) 
that $\Gamma^c$ is $\omega$-categorical as well.

We want to show that $\Gamma^c$ is a model-complete core.
Suppose that $e$ is an endomorphism of $\Gamma^c$. 
We have to show that $e$ is an elementary embedding.
Let $\phi$ be a first-order formula. Then $\phi$ is in $\Delta^c$ 
equivalent to a quantifier-free formula $\psi$.
Note that $e$ preserves existential positive formulas, and because
$\Delta^c$ is a core, $e$ preserves $\psi$ as well. Therefore,
$e$ is elementary.

Now suppose that $\Gamma'$ is some model-complete core of $\Gamma$.
We show that $\Gamma'$ is isomorphic to $\Gamma^c$.
In fact, we will show that $\Delta^c$ and the expansion $\Delta'$ 
of $\Gamma'$ 
by all existential positive definable relations have the same age and
are both homogeneous. Then Theorem~\ref{thm:fraisse} implies that
$\Delta^c$ and $\Delta'$, and hence also $\Gamma^c$ and $\Gamma'$
are isomorphic. 

Similarly as in the proof of Proposition~\ref{prop:premain} 
we prove that every existential formula that holds
on tuples from $\Delta'$ in $\Delta$ also hold in
$\Delta'$. This also shows that $\Delta'$ is $\omega$-categorical.
To show that $\Delta'$ is homogeneous, by Proposition~\ref{prop:qe}
it suffices to show that $\Delta'$ has quantifier elimination.
Let $\phi$ be a first-order formula. Since $\Gamma'$ is model-complete,
Proposition~\ref{prop:mc} shows that $\phi$ is in $\Gamma'$ and therefore
also in $\Delta'$ equivalent
to an existential formula. Since $\Delta'$ is $\omega$-categorical and
a core, we can then apply Lemma~\ref{lem:exposex} to show that 
$\phi$ is equivalent to an existential positive formula in $\Delta'$.
Since $\Delta'$ contains all existential positive formulas,
we have shown that $\Delta'$ admits quantifier elimination.

Finally, let $R$ be an orbit $R$ of $k$-tuples in the automorphism 
group of $\Gamma^c$. All $k$-tuples in $R$ 
induce isomorphic substructures $S$ in $\Delta^c$. 
Because $\Delta^c$ is homogeneous,
all $k$-tuples in $\Delta^c$ that are isomorphic to $S$ 
are contained in $R$. 
Thus, $R$ has a definition as a conjunction $\varphi$ of atomic formulas.
We replace all relation symbols in $\varphi$ that are contained in 
the signature of $\Delta^c$ but not in the signature of $\Gamma^c$ by
their existential positive definition. Then the resulting
formula can be re-written as a disjunction of primitive positive formulas.
Every disjunct is either false or already defines $R$, because
$R$ is an orbit of $k$-tuples in Aut$(\Gamma^c)$. Hence, the formula
is equivalent to a single disjunct, and $R$ is pp-definable in $\Gamma^c$.
\end{proof}

\begin{corollary}
If $\Gamma$ is an $\omega$-categorical structure that is
expanded by all pp-definable relations.
Then $\Gamma$ is homomorphically equivalent to a
homogeneous core $\Gamma^c$ (which is again unique up to isomorphism).
\end{corollary}
\begin{proof}
Let $\Gamma^c$ be the $\omega$-categorical model-complete core that is
homomorphically equivalent to $\Gamma$. Theorem~\ref{thm:main}
shows that every orbit of $k$-tuples in the automorphism group
of $\Gamma^c$ has a pp-definition in $\Gamma^c$,
and the proof of Theorem~\ref{thm:main} shows that 
because $\Gamma$ is expanded by all pp-definable
relations, $\Gamma^c$ contains all pp-definable
relations as well. We have also seen that 
in $\Gamma^c$, every first-order definable
relation has a quantifier-free definition in the expanded 
structure that contains all existential positive formulas.
Every existential positive formula is a disjunction
of primitive positive formulas, and therefore the structure $\Gamma^c$ has
quantifier-elimination as well. Proposition~\ref{prop:qe}
implies that $\Gamma^c$ is homogeneous.
\end{proof}

We also obtain alternative characterizations of when a model-complete
$\omega$-categorical structure is a core.
A set of functions $F$ from $\Gamma$ to $\Gamma$ \emph{locally generates}
a function $g$, if for every finite subset $A$ of $\Gamma$ there is a function
$f\in F$ such that $g(a)=f(a)$ for all $a\in A$
(this is a standard notion in universal algebra~\cite{Szendrei}).

\begin{theorem}\label{thm:mc}
Let $\Gamma$ be a model-complete $\omega$-categorical structure. 
Then the following are equivalent.
\begin{enumerate}[(1)]
\item $\Gamma$ is a core.
\item Every first-order formula 
is in $\Gamma$ equivalent to an existential positive formula.
\item Every endomorphism of $\Gamma$ is an elementary embedding.
\item The automorphism group of $\Gamma$ locally generates the 
endomorphism monoid of $\Gamma$. 
\end{enumerate}
\end{theorem}

\begin{proof}
Every first-order definable $k$-ary
relation $R$ in $\Gamma$ 
is the union of a finite number of orbits of $k$-tuples of Aut$(\Gamma)$.
Theorem~\ref{thm:main} 
shows that each of these orbits has a pp-definition. 
A disjunction of primitive positive formulas
can be equivalently written as an existential positive formula,
and we obtain an existential positive definition for $R$,
showing that 1 implies 2.

Since endomorphisms clearly preserve existential positive formulas,
2 implies that every endomorphism is elementary.

A function from $\Gamma$ to $\Gamma$
is in the local closure of the automorphism group 
of $\Gamma$ if and only
if it preserves all first-order definable relations.
If we assume that every endomorphism is elementary, then 
the automorphism group locally generates all endomorphisms of $\Gamma$,
and therefore 3 implies 4.

To prove that 4 implies 1, assume that $\Gamma$ has
a strict endomorphism $f$, i.e., there is a tuple $(u_1, \dots, u_k)$ 
in $\Gamma$ such that
$f(u_1) = f(u_2)$, or $R(f(u_1), \dots, f(u_k))$ and not $R(u_1, \dots, u_k)$.
Clearly, such a function $f$ cannot be locally generated by automorphisms.
\end{proof}

%% file: constants.tex
\section{Adding Constants to the Signature}\label{sect:constants}


In relational structures $\Gamma$ we can use singleton relations, i.e., relations of 
the form $\{c\}$ for an element $c$ from the domain of $\Gamma$, to model
the concept of \emph{constants} in first-order structures~\cite{Hodges}. One of the main results in~\cite{JBK} says
that if $\Gamma$ is a finite core, then adding a singleton-relation
does not increase the complexity of the corresponding 
constraint satisfaction problem.
We show that the same holds for constraint satisfaction problems 
where the template is an $\omega$-categorical model-complete core.
Note that this directly applies to all the computational problems presented
in the introduction and Sections~\ref{sect:cat} and~\ref{sect:hom}.

\begin{theorem}\label{thm:constants}
Let $\Gamma$ be a model-complete $\omega$-categorical core,
and let $\Gamma'$ be the expansion of $\Gamma$ 
by a unary singleton relation $C = \{c\}$.
If $\text{CSP}(\Gamma)$ is tractable, then so is $\text{CSP}(\Gamma')$.
(If $\text{CSP}(\Gamma')$ is NP-hard, then so is $\text{CSP}(\Gamma)$.)
\end{theorem}
\begin{proof}
We show how to solve CSP$(\Gamma')$ in polynomial time, 
under the assumption
that CSP$(\Gamma)$ can be solved in polynomial time.
Let $S'$ be an instance of CSP$(\Gamma')$. Let $P$ be the orbit of $c$ 
in the automorphism group of $\Gamma$. By Theorem~\ref{thm:main}, 
$P$ is primitive positive definable in $\Gamma$. 
Thus we can assume without loss of generality that 
$\Gamma$ and $\Gamma'$ contain the relation $P$.
Replace all occurrences of the relation $C$ in $S'$ 
by the relation $P$. Solve the resulting instance $S$ of CSP$(\Gamma)$;
by assumption this is possible in polynomial time.
If $S$ is not satisfiable, then in particular $S'$ was not satisfiable. 
On the other hand, if there is a homomorphism $h$ from
$S$ to $\Gamma$, we claim that there is a homomorphism from $S'$ to $\Gamma'$.
Since $P$ is the orbit of the element $c$, there is an automorphism $\alpha$ 
of $\Gamma$ such that $h \alpha$ is a solution of the instance $S'$ 
of CSP$(\Gamma')$.
\end{proof}

If an $\omega$-categorical model-complete 
core $\Gamma$ is expanded by a singleton relation,
the resulting structure $\Gamma'$ is again an $\omega$-categorical model-complete core. The fact that $\omega$-categoricity is preserved by
such expansions is well-known, 
see e.g.~\cite{Hodges}. Suppose that $\Gamma'$ is not a model-complete 
core, i.e., 
there is a non-elementary endomorphism $e$ of $\Gamma'$. 
Then $e$ is also a non-elementary endomorphism
of $\Gamma$, a contradiction.
Hence, we can apply Theorem~\ref{thm:constants} several times, and
obtain that Theorem~\ref{thm:constants} also holds for expansions
by a \emph{finite} number of singleton relations.